\documentclass[twocolumn,epjc3]{svjour3} % use epjc3 option for EPJC

\smartqed

\RequirePackage{graphicx}
\RequirePackage{amsmath,amssymb,amsfonts}
\RequirePackage[numbers,sort&compress]{natbib} % numeric, compress citations
\usepackage{orcidlink}      % then load orcidlink
\usepackage{keyval,amsfonts, slashed, bm}
\usepackage{graphicx,textcomp}
\usepackage{epsfig,amssymb}
\usepackage{subfig,cancel}
\usepackage{multirow}% http://ctan.org/pkg/multirow
\usepackage{MnSymbol}
\usepackage{graphicx}% Include figure files
\usepackage[font=small,labelfont=bf,justification=justified]{caption}
\usepackage{color}
\definecolor{blue(munsell)}{rgb}{0.2, 0.3, 0.69}
\definecolor{coquelicot}{rgb}{1.0, 0.22, 0.0}
\newcommand{\ba}{\begin{eqnarray}}
\newcommand{\ea}{\end{eqnarray}}

\newcommand{\nn}{\nonumber\\}

\definecolor{sinopia}{rgb}{0.8,0.25,0.04}

\definecolor{greenopia}{rgb}{0.3,0.65,0.14}

\makeatletter
\newsavebox{\@brx}
\usepackage[usenames,dvipsnames]{xcolor}
\usepackage{color}
\usepackage[english]{babel}
\usepackage{dcolumn}
\usepackage{pifont}
\usepackage{dsfont,mathrsfs}
\usepackage{cancel}
\usepackage{bigints}
\usepackage{accents}
\usepackage{soul}
\usepackage{multirow}
\usepackage{natbib}
\allowdisplaybreaks

\usepackage{lineno}
\def\nn {\nonumber}

\def\be {\begin{equation}}
	\def\ee {\end{equation}}
\def\bea {\begin{eqnarray}}
	\def\eea {\end{eqnarray}}
\def\bc {\begin{center}}
	\def\ec {\end{center}}

\journalname{Eur. Phys. J. C}



\begin{document}

\title{Electrical conductivity of QGP with quasiparticle quarks and Gribov gluon}                      
\author{Sadaf Madni \thanksref{1}    \orcidlink{0000-0002-8251-2856}  
\and Sumit  \thanksref{2} \orcidlink{0000-0001-7137-6433}
\and Lata Thakur    \thanksref{3,4}  \orcidlink{0000-0003-2343-4963} 
\and Najmul Haque \thanksref{1}  \orcidlink{0000-0001-6448-089X} 
} 

\institute{\label{1} School of Physical Sciences, National Institute of Science Education and Research, An OCC of Homi Bhabha National Institute, Jatni-752050, India 
\and \label{2} School of Physics, Beijing Institute of Technology, 102488 Beijing, China  
\and \label{3} Department of Physics and Institute of Physics and Applied Physics, Yonsei University, Seoul 03722, Korea 
\and \label{4} Asia Pacific Center for Theoretical Physics, Pohang, Gyeongbuk 37673, Republic of Korea 
}

\thankstext{e1}{\email{sadaf.madni@niser.ac.in}}
\thankstext{e3}{\email{sumit@ph.iitr.ac.in}}
\thankstext{e2}{\email{thakurphyom@gmail.com}}
\thankstext{e4}{\email{nhaque@niser.ac.in}}

\date{Received: date / Accepted: date}

\maketitle
	
\begin{abstract}
We investigate the electrical conductivity of the quark–gluon plasma (QGP) using a non-perturbative resummation scheme incorporating the Gribov-modified gluon propagator. The electrical conductivity is evaluated by solving the relativistic Boltzmann transport equation within the relaxation-time approximation, where the relaxation times are obtained from microscopic two-body scattering amplitudes. A quasiparticle description is employed for quarks, providing a unified framework for studying transport properties across both weakly and strongly coupled regimes. Above the deconfinement transition temperature, we estimate the electrical conductivity of the QGP and compare our results with available lattice QCD data and various phenomenological models, finding good agreement with the lattice results.
\end{abstract}	

\maketitle
%%%%%%%%%%%%%%%%%%%%%%%%%%%%%%%%%%%%%%%%%%%%%%%%%
%\medskip
%%%%%%%%%%%%%%%%%%%%%%%%%%%%%%%%%%%%%%%%%%%%%%%%%	

\section{Introduction}
The discovery of the quark–gluon plasma (QGP) at the beginning of this century has opened new directions in the study of strongly interacting matter through relativistic heavy-ion collision (HIC) experiments at RHIC and the LHC~\cite{Gyulassy:2004zy, Jacobs:2004qv, Busza:2018rrf}. One of the principal objectives of these experimental programs is to obtain precise estimates of the transport properties of the QGP. Over the past two decades, extensive theoretical and phenomenological studies employing ideal~\cite{Teaney:2000cw, Huovinen:2001cy, Hirano:2002ds, Broniowski:2008vp, Schenke:2010nt}, viscous hydrodynamics~\cite{Romatschke:2007mq, Song:2007fn, Dusling:2007gi, Bozek:2009dw, Bozek:2012qs, Ryu:2015vwa, Du:2019obx} have revealed that the QGP is a strongly coupled system behaving as an almost perfect fluid in HICs. The hydrodynamic description of QGP provides valuable insights into its evolution and transport characteristics~\cite{Meyer:2007dy, Heinz:2013th, Jeon:2015dfa}. 

Transport properties, characterized by the corresponding transport coefficients, encode essential information about microscopic interactions within the medium and serve as crucial theoretical inputs for the hydrodynamic modeling of QGP. These coefficients are indispensable for analyzing HIC data and understanding the dynamical properties of QGP~\cite{Gale:2013da, Schenke:2011zz, Heinz:2013th}. 

In relativistic HIC, extremely strong electromagnetic fields are also generated~\cite{Tuchin:2013ie}. The influence of these fields on QGP dynamics, however, remains a topic of active debate. Some studies suggest that the fields decay rapidly with time and thus have limited impact on the medium’s evolution, while others argue that a finite electrical conductivity can significantly prolong their lifetime, potentially leading to observable effects. Once local equilibrium is established, the electrical conductivity ($\sigma_\text{el}$) of the medium becomes a key parameter governing how electromagnetic fields evolve and interact with the plasma.
The electrical conductivity of QGP has been extensively studied within various theoretical frameworks~\cite{Arnold:2000dr, Arnold:2003zc, Gupta:2003zh, Aarts:2007wj, Buividovich:2010tn, Ding:2010ga, Burnier:2012ts, Brandt:2012jc, Amato:2013naa, Aarts:2014nba, Cassing:2013iz, Steinert:2013fza, Hirono:2012rt, Greif:2014oia, Puglisi:2014pda, Puglisi:2014sha, Finazzo:2013efa, Greif:2016skc, Mitra:2016zdw, Srivastava:2015via, Thakur:2017hfc, Marty:2013ita, Kadam:2017iaz, Thakur:2019bnf, Berrehrah:2016vzw, Mitra:2018akk, Soloveva:2019xph, Singha:2017jmq, Mitra:2017sjo, Ghosh:2024fkg}. During the early stage of the collision, electric currents are generated by the quarks, with $ \sigma_\text{el} $ governing this production process. The value of $ \sigma_\text{el} $ plays a fundamental role in determining the strength of the chiral magnetic effect~\cite{Fukushima:2008xe}, which is a signature of CP violation in strong interactions~\cite{Kharzeev:2007jp}. In mass asymmetric collisions (such as Cu-Au collisions), the electrical field has a preferred direction, which generates a charge asymmetric flow~\cite{Hirono:2012rt}. The strength of this flow is directly related to $ \sigma_\text{el} $, which is associated with the emission rate of soft photons~\cite{Turbide:2003si, Linnyk:2013wma}. 
%The electrical conductivity of the QCD medium has gained attention due to a strong electric field created in the collision zone of ultra-relativistic heavy-ion collision experiments. 
%In peripheral heavy-ion collisions, a large electrical and magnetic field is generated, which can significantly impact the medium's behavior~\cite{Tuchin:2013ie}. 
%%The behavior of the medium formed during collisions can be significantly affected by a large electrical field. The magnitude of electrical conductivity ($ \sigma_{el} $) of the medium plays a crucial role in determining the effect of the electrical field. 

In this work, we compute the electrical conductivity of the QGP medium %with quasiparticle quarks and Gribov gluons 
using the Gribov-Zwanziger (GZ) approach~\cite{Gribov:1977wm, Zwanziger:1989mf}. This approach has gained significant attention, %in the heavy-ion theory community, 
particularly after its generalization to finite temperature QCD medium~\cite{Zwanziger:2004np, Zwanziger:2006sc}. Studies have shown that the infrared mass scale parameter, $ \gamma_G $ which is an intrinsic Yang-Mills scale, originally introduced by Gribov to explain the non-perturbative confinement region, significantly improves the infrared behavior of QCD and leads to a good agreement with lattice results for thermodynamic quantities~\cite{Fukushima:2013xsa} along with the novel massless excitations ascribable to the magnetic scale~\cite{Su:2014rma}. The Gribov dispersion relation provides a simple way to account for the effects of residual confinement on the transport properties of the QGP. It was first used in Refs.~\cite{Florkowski:2015dmm, Florkowski:2015rua, Begun:2016lgx} in the context of kinetic theory and hydrodynamics, where it was applied to a boost-invariant setup. Also, its impact on observables, such as the dilepton rate and quark number susceptibility, has been examined~\cite{Bandyopadhyay:2015wua}. Additionally, the shear and bulk viscosity of the QGP medium have been explored using Gribov gluons and quasiparticle quarks~\cite{Madni:2024xyj} along with mesonic screening masses~\cite{Sumit:2023hjj}, spectral sum rules for quarks~\cite{Du:2024sbv}, and the different properties of heavy quark sector~\cite{Debnath:2023dhs, Madni:2022bea, Sumit:2023oib, Sumit:2025ddb, Mazumder:2024grc}, have also been investigated using the GZ scheme. Recently, a covariant kinetic theory has been developed to study the transport coefficients for Gribov plasma~\cite{Jaiswal:2020qmj}. The study utilized a quasiparticle-like framework with bag correction for pressure and energy density. 
The temperature dependence of the Gribov parameter was obtained from pure gauge lattice thermodynamics~\cite{Jaiswal:2020qmj}, whereas the running coupling was fixed using the $(2+1)$-flavor lattice QCD equation of state.
%The temperature dependence of the Gribov parameter and running coupling was determined by matching with lattice results for a system of gluons. 
We have utilized these parameters in this work to calculate the electrical conductivity of the QGP medium.
%Lattice calculations have shown inconsistencies in transport coefficient results, making it necessary to find alternate methods to incorporate non-perturbative features in the theory. %We follow the approach suggested by Gribov in his seminal work~\cite{Gribov:1978} to estimate the electrical conductivity of the QGP medium. 

This article is divided into six sections. Section~\ref{formalism} discusses the formalism of the current work in detail, where the Gribov parameter and running coupling, $g$, are defined. Section~\ref{crosssection} focuses on the scattering cross-section of the constituent particles and computes the scattering amplitude of quark-quark and quark-antiquark interaction. In Section~\ref{relaxationtime}, the relaxation time is revisited, which is based on the weighted thermally averaged quark-quark and quark-antiquark cross-section. Section~\ref{electricalconductivity} investigates the electrical conductivity of the medium by utilizing the quasiparticle model. Finally, Section~\ref{summary} provides an overview and outlook of the work.
%%%%%%%%%%%%%%%%%%%%%%%%%%%%%%%%%%%%%%%%%%%%%%%%%
\vspace{-.5cm}
\section{Formalism}\label{formalism}
	
\subsection{Gribov parameter and running coupling}
%		\subsection{Gribov-parameter, $\gamma_G$}
The Gribov-modified gluon propagator, in the Landau gauge, reads~\cite{Su:2014rma}
\begin{align}
\mathcal{D}_{ab}^{\mu\nu}(k)=\delta_{ab}\frac{k^2}{k^4+\gamma_G^4}\left(\delta^{\mu\nu}-\frac{k^{\mu}k^{\nu}}{k^2}\right)~,
\label{eq:gribovprop}
	\end{align}
where $k^\mu$ is the Euclidean gluon four momentum, and the temperature dependence of the Gribov parameter is determined from pure gauge lattice thermodynamics~\cite{Jaiswal:2020qmj}. Note that the Gribov parametrization has been extensively studied in the literature to explain deconfined nuclear matter~\cite{Zwanziger:2004np, Zwanziger:2006sc, Fukushima:2013xsa, Florkowski:2015dmm, Florkowski:2015rua, Begun:2016lgx, Bandyopadhyay:2015wua, Su:2014rma}. 
%Ref.~\cite{Zwanziger:2006sc} justifies the use of the Gribov prescription to describe deconfined nuclear matter and explains in detail that the evidence for confinement was observed from lattice simulations.
	
It is important to realize that in this work, we have employed the general form of the Gribov-modified gluon propagator (Eq.~\eqref{eq:gribovprop}) to evaluate the electrical conductivity of the QGP medium. For simplicity and to focus on the primary effects of the Gribov modification, we have neglected any contributions from gluon condensate terms in the propagator, which would lead to the emergence of a dynamical mass term(see~Ref.\cite{Dudal:2008sp}). It is important to note that recent lattice studies indicate a preference for the so-called decoupling solution, where the gluon propagator remains finite in the limit \(k\rightarrow 0\). This behavior can be incorporated into a more refined version of the GZ framework that includes such condensate effects, leading to a modified propagator with a dynamical mass term. However, incorporating these refinements introduces additional complexities and model dependencies that are beyond the scope of the present work. Here, we opt to use the original Gribov form of the gluon propagator without these condensate-induced modifications. This choice allows us to isolate the impact of the Gribov parameter on the electrical conductivity. It provides a clear and straightforward framework for understanding its effects on the QGP's transport properties. While this approach represents a simplified treatment, it serves as a baseline study to elucidate the key features of Gribov-modified gluons in QGP without the additional complications arising from dynamical mass terms.

To characterize the system's dynamics, the energy-momentum tensor takes on the following form~\cite{Jaiswal:2020qmj, Jeon:1995zm}
	\begin{align}
		T_{(0)}^{\mu\nu}= \int dk \cdot k^{\mu}k^{\nu} {\rm{g}}^0+B_0(T) g^{\mu\nu}~,
		\label{Tmunu}
	\end{align}
	where $B_0(T)$ is bag pressure, which is added to take care of thermodynamic consistency in equilibrium, $g^{\mu\nu}$ is the metric tensor where \mbox{$g^{\mu\nu}={\rm diag}(+1,-1,-1,-1)$} and ${\rm{g}}^0$ is the equilibrium distribution function for Gribov plasma, with ${\rm g}^{0}_{\pm}$ corresponding to the two dispersion branches, defined as,
	\begin{equation}
		  {\rm{g}}_{\pm}^{0}=\frac{1}{e^{\epsilon_{\pm}/T}-1}, 
		\label{g0}
	\end{equation}
	where $\epsilon_{\pm}=\sqrt{|\bm{k}|^2\pm i\gamma_G^2}$ with $|\bm{k}|$ as the three-momentum of the gluons. The Lorentz invariant momentum integral is given by
	\begin{align}
		\int dk=\frac{d_g}{(2\pi)^3}\int d^3\bm{k} \int dk_0 \hspace{0.1cm}  2 \times  \Theta(k_0) \hspace{0.1cm} \delta\left(k^2+\frac{\gamma_G^4}{k^2}\right).
		\label{dpgluon}
	\end{align}
	%%%
In the above equation, the degeneracy factor, $d_g=2\times (N_c^2-1)=16$. Note that in the Landau gauge, the Gribov framework also comes with ghost degrees of freedom. At very large temperatures, these ghost degrees of freedom cancel a longitudinal gluon mode that contributes to the expression of the free-energy as well as one of the three transverse gluon modes present in Eq.~\eqref{eq:gribovprop}. At intermediate to low temperatures, however, it is actually not known whether this cancellation occurs within the Gribov framework. Unlike the conventional Gribov formalism, this work accounts for medium effects through the Gribov parameter, keeping the number of gluon modes unchanged. This is done on the same footing as HTL perturbation theory, where the medium effect is incorporated via the Debye mass $m_D$~\cite{Ghiglieri:2020dpq,Haque:2024gva}. Hence, we are considering %$d_g=2$, counting 
only two transverse polarizations, and we are deviating from the standard Gribov framework. 

Now, the equilibrium pressure and energy density can be obtained from Eq.~\eqref{Tmunu} using the relation
\begin{equation} \label{pressure1}
\mathcal{P}_{\rm eq} = -\frac{1}{3}\Delta_{\mu\nu}T_{(0)}^{\mu\nu}=\mathcal{P}_{\rm GZ}-B_0(T)~. 
\end{equation}
\begin{equation} \label{energy1}
		\mathcal{E}_{\rm eq} = u_{\mu}u_{\nu}T_{(0)}^{\mu\nu}=\mathcal{E}_{\rm GZ}+B_0(T)~.
\end{equation}
where $\Delta_{\mu\nu}=g_{\mu\nu}-u_{\mu}u_{\nu}$, $u_{\mu}$ is the fluid four velocities satisfying $u^{\mu}u_{\mu}=1$. In fluid rest frame, $u^{\mu}=(1,0,0,0)=(1,\vec{0})$. $\mathcal{P}_{\rm GZ}$ and $\mathcal{E}_{\rm GZ}$ are the particle contributions to pressure and energy density of equilibrium Gribov plasma, which  gives 
\begin{align}
		\mathcal{P}_{\rm GZ}=\frac{d_g}{(2 \pi)^3} \int \mathrm{d}^3 \bm{k} \frac{|\bm{k}|^2}{6}\left(\frac{{\rm{g}}^0_{+}}{\epsilon_{+}}+\frac{{\rm{g}}^0_{-}}{\epsilon_{-}}\right), \\
		\mathcal{E}_{\rm GZ}=\frac{d_g}{(2 \pi)^3} \int \mathrm{d}^3 \bm{k} \frac{1}{2}\left({\rm{g}}^0_{+} \epsilon_{+}+{\rm{g}}^0_{-} \epsilon_{-}\right).
\end{align}
The entropy density of Gribov plasma can be calculated using Eqs.~\eqref{pressure1} and~\eqref{energy1} by utilizing the thermodynamic relation
\begin{align}
s_{\rm GZ}=\frac{\mathcal{P}_{\rm GZ}+\mathcal{E}_{\rm GZ}}{T}= \frac{\mathcal{P}_{\rm eq}+\mathcal{E}_{\rm eq}}{T}= s_{\rm eq}.	
		\label{entropygribov}
\end{align}
where $s_{\rm eq}$ denotes the entropy density (for a pure gauge theory) obtained within the Gribov framework. To fix the Gribov parameter, the first step is to match the temperature dependence of the scaled trace anomaly of pure gauge lattice results~\cite{Borsanyi:2012ve}, in order to fix the equilibrium thermodynamic quantities. 
	For analytical tractability, the trace anomaly is fitted with a specific functional form~\cite{Borsanyi:2010cj}
	\begin{align}
		\frac{\mathcal{I}_{\rm eq}}{T^4}=&\frac{\mathcal{I}_{\rm lat}^F}{T^4}= \exp\left[-\left(\frac{a_1}{\Tilde{T}}+\frac{a_2}{\Tilde{T}^2}\right)\right]\nonumber\\
		&\times \left(\frac{a_0}{1+a_3 {\Tilde{T}}^2}+\frac{b_0({\tanh}[b_1\Tilde{T}+b_2]+1)}{1+c_1\Tilde{T}+c_2\Tilde{T}^2} \right),
		\label{TAPG}
	\end{align}	
	where $\Tilde{T}$ is the scaled temperature ($=T/T_c$). 
	
	\begin{table}[tbh]
		\centering
		\begin{tabular}{|c|c|c|c|c|c|c|c|c|}
			\hline
			$a_0$ & $a_1$  & $a_2$  & $a_3$  & $b_0$ & $b_1$ & $b_2$ & $c_1$ & $c_2$ \\
			\hline
			$0.23$  & $-1.83$ & $2.92$ & $0.07$ & $0.32$ & $62.39$ & $-62.55$ & $-1.98$ & $1.08$\\
			\hline
		\end{tabular}
		\caption{The parameters extracted from the function (Eq.~\eqref{TAPG}) after fitting with the pure gauge lattice data.} 
	\end{table}
	
	For this set of parameters, the scaled pressure can be obtained by
	\begin{align}
		\frac{\mathcal{P}_{\rm eq}(T)}{T^4}=\frac{\mathcal{P}_{\rm lat}(T)}{T^4}=\frac{\mathcal{P}_{\rm lat}(T_{\rm eq})}{T_{eq}^4}+\int_{T_{\rm eq}}^{T}\frac{d\tilde{T}}{\tilde{T}}\times \frac{\mathcal{I}_{\rm lat}^F}{\tilde{T}^4} ~,
		\label{fittedpressuregluon}
	\end{align} 
where $\mathcal{P}_{\rm lat}/T_{\rm eq}^4=0.0015$ and $T_{\rm eq}=0.7\,T_c$. Following the Eqs.~\eqref{TAPG} and \eqref{fittedpressuregluon}, $\mathcal{I}_{\rm lat}=  \mathcal{E}_{\rm lat}-3\mathcal{P}_{\rm lat}$ provides the variation of energy density, $\mathcal{E}_{\rm lat}$ as a function of scaled temperature. To compute the lattice entropy density $s_{\rm lat}$, we use the thermodynamic relation, $s_{\rm lat}=d\mathcal{P}_{\rm lat}/dT=(\mathcal{P}_{\rm lat}+\mathcal{E}_{\rm lat})/T$. Therefore, using Eq.~\eqref{entropygribov} for the entropy density of the Gribov-modified gluon, we can equate $ s_{\rm GZ}=s_{\rm lat}$, which yields the fixed values of the Gribov parameter as follows in Ref.~\cite{Jaiswal:2020qmj}.

Note that for the determination of the Gribov parameter, the thermodynamic matching is performed using pure gauge lattice results, while for the electrical conductivity calculation, we incorporate the quark degrees of freedom in the quasiparticle description, effectively extending the framework towards phenomenological full QCD observables. Thus, the total entropy density of the QGP medium is the sum of the entropy densities of the Gribov-modified gluons and the quasiparticle quarks, i.e
\begin{align}
		s^{\rm QGP}_{\rm total}=s_{\rm GZ(g)}+s_{\rm qp(q)}~,
	\label{enttotal}
	\end{align}
	where `qp(q)' represents the quasi-particle quark and entropy for qp(q) can be obtained as
		\begin{align}
		s_{\rm qp(q)}=\sum_{i=q,\bar{q}}\frac{d_i}{2\pi^2}\int_{0}^{\infty}\bm{k}^2 d\bm{k} \frac{(\frac{4}{3}\bm{k}^2+m_i^2)}{E_i T}f_i^0~.
		\label{s_qp}
	\end{align}
%%%%%%
In Eq.~\eqref{s_qp}, the index $i$ runs over quarks and antiquarks, i.e., $i=q,\bar q$. Here the light quark sector consists of $u$ and $d$ flavors, while $s$ denotes the strange quark. The degeneracy factor $d_i$ accounts for spin and color degrees of freedom and is given by $d_{q_l,\bar q_l}=2\times N_c\times N_{q_l}$ for light quarks (anti-quarks) and $d_{s,\bar s}=2\times N_c$ for strange quarks (anti-quarks), where $N_c=3$ and $N_{q_l}=2$ correspond to the number of colors and light quark flavors, respectively.
%In Eq.~\eqref{s_qp}, $i (= q, \bar{q})$ is the sum over light and strange (anti-) quarks. $d_i$ is the degeneracy factor for spin and color; which reads $d_{q_l,\bar{q_l}}=2\times N_c\times N_{q_l,\bar{q_l}}$ for light quarks (anti-quarks), $d_{s,\bar{s}}=2\times N_c$ for strange quarks with $N_c=3$ and $N_{q_l}=2$. Here, $f_i^0$ is the equilibrium distribution for fermions, which is defined as
	\begin{equation}
		f_i^{0}\equiv f_i^{0}(E_i,T)=\frac{1}{e^{E_i/T}+ 1}, 
		\label{f0}
	\end{equation} 
where $E_i=\sqrt{|\bm{k}|^2+m^2_i}$ is the quasi-particles energies in thermal equilibrium with $m_i$ as the effective mass. The effective mass depends on the temperature and chemical potential, which arises due to the interaction of quarks and gluons with the surrounding matter and is given by~\cite{Mykhaylova:2019wci}
\begin{equation}
m^2_i=m^2_{i0}+\rm{\Pi}_i,
\label{m_eff}
\end{equation}
	where $m_{i0}$ is the bare mass and $\rm{\Pi}_i$ is the dynamically generated self-energy, which can be obtained by using the hard-thermal loop approximation (HTL) in asymptotic forms as
	\begin{align}
	& {\rm\Pi}_{q_l}(T)=\frac{g^2T^2}{3} \left(m_{q_l0} \frac{\sqrt{6}}{gT} +1\right), \nonumber \\
	& {\rm\Pi}_s(T)=\frac{g^2T^2}{3}\left(m_{s0} \frac{\sqrt{6}}{gT} + 1 \right).
	\label{quasimass}
	\end{align}
	where $m_{q_l0}$ and $m_{s0}$ are the bare masses of the light  ($u,d$) and strange ($s$) quarks, whose values are taken as $5$ MeV and $95$ MeV respectively. Here, we limit our study to the medium at finite temperature and vanishing chemical potential. We are considering the medium, which consists of $u$, $d$, and $s$ quarks, which interact via the exchange of Gribov-modified gluons. 
	
	\begin{figure}
		\centering
		\includegraphics[scale=0.54]{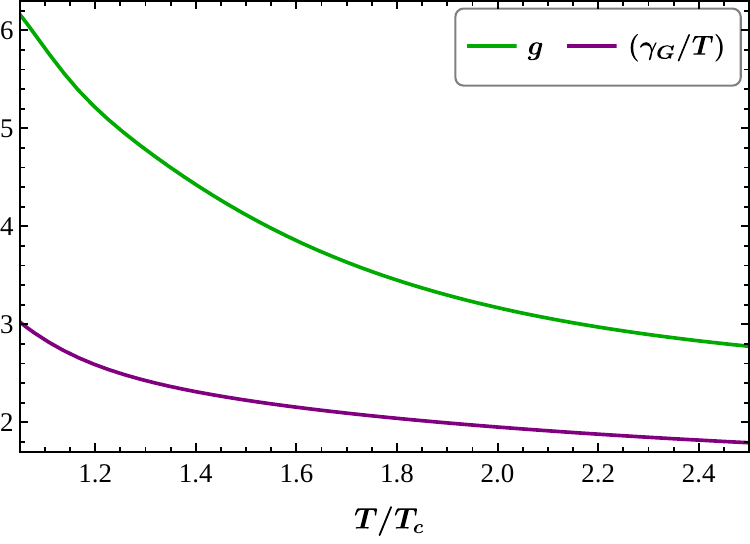}
		\includegraphics[scale=0.55]{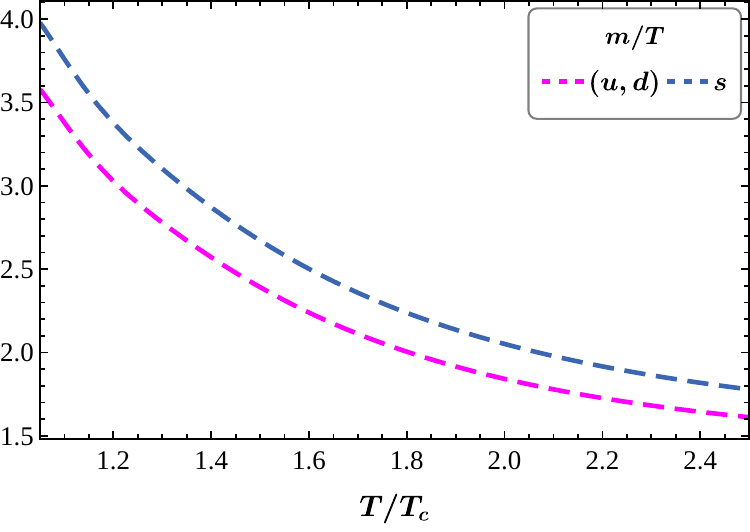}
                     %\caption{The running coupling $g(T)$ as a function of $T/T_c$ fitted using the lattice data of the equation of state of (2+1) QCD from Ref.~\cite{Borsanyi:2013bia} and the scaled Gribov parameter {\rd{$\gamma_G/T$, obtained from pure gauge lattice thermodynamics}}. Also, the quasiparticle masses(Eq.\eqref{quasimass}) of the quarks, scaled with the temperature, are shown in the plot against the scaled temperature in the interval $1.05 \le T/T_c \le 2.5$.} %The critical temperature($T_c$) for the medium QGP has been taken as 0.158 GeV. }
\caption{The running coupling $g(T)$ as a function of $T/T_c$, fitted using the lattice equation of state of $(2+1)$-flavor QCD from Ref.~\cite{Borsanyi:2013bia}. Also shown are the scaled Gribov parameter $\gamma_G/T$, obtained from pure gauge lattice thermodynamics, and the quark quasiparticle masses (Eq.~\eqref{quasimass}), scaled by temperature, plotted as functions of $T/T_c$ in the interval $1.05 \le T/T_c \le 2.5$.}
	\label{fig:coupling}
	\end{figure}
    
	The interactions of quarks and gluons with the surrounding matter in the medium are encoded in the quasiparticle masses (Eq.~\ref{m_eff}), which depend on the running coupling, $g$. Here, we fix the running coupling using the lattice data of entropy density for (2+1) QCD. The fit function for the entropy density for the medium consisting of the quasiparticle quarks has been formulated much like for the gluonic ($N_f=0$) case. Therefore, to fix the running coupling, we equated the total entropy density of the QGP medium with the entropy density from the lattice, see Eq.~\eqref{enttotal}.

	In Fig.~[\ref{fig:coupling}], we plot the running coupling, $g(T)$, and the scaled Gribov parameter as a function of scaled temperature ($T/T_c$). 
	%The data were fitted using the lattice equation of the state of (2+1) QCD. 
    The critical temperature, $T_c$, is taken to be 0.155 GeV. The green line represents the running coupling, and the purple line represents the scaled Gribov parameter. We find that both the Gribov parameter and running coupling decrease monotonically with an increase in temperature above $T_c$. 
	%%%%%%%%%%%%%%%%%%%%%%%%%%%%%%%%%%%%%%%%%%%%%%%%%%%%%%%%%%%%%%%%%%%%%%%%%%%%%%%%%%%%%%%%%%%%%%%%%%
\section{Elastic cross-sections}\label{crosssection}	
\begin{figure}
	\centering
	\includegraphics[width=\linewidth]{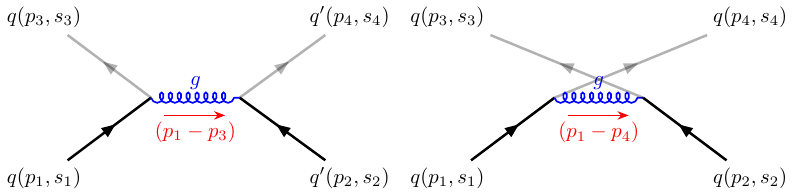}
	\caption{Feynman diagram for $qq^{\prime} \rightarrow qq^{\prime}$ processes. Left: $t-channel$ and Right: $u-channel$, when $q^{\prime}=q$. The black line corresponds to the incoming quarks, while the light grey corresponds to the outgoing quarks. The ($p_i,s_i$) is the four-momentum and the spin of the considered quark.}
	\label{qqqq}
\end{figure}

\begin{figure}
	\centering
	\includegraphics[width=\linewidth]{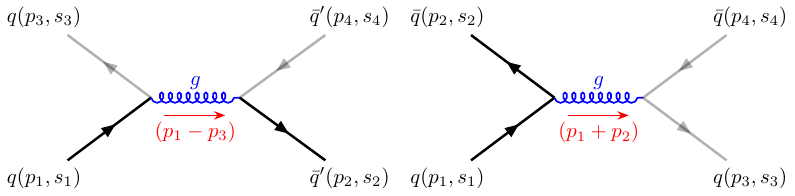}
	\caption{Feynman diagram of $q\bar{q}^{\prime} \rightarrow q\bar{q}^{\prime}$. Left: $t-channel$ and Right: $s-channel$, when $\bar{q}^{\prime}=\bar{q}$. The black line corresponds to the incoming quark/antiquark, while the light grey corresponds to the outgoing quark/antiquark. The ($p_i,s_i$) is the four-momentum and the spin of the considered (anti)quark.}
	\label{qqbarqqbar}
\end{figure}

To analyze the transport properties of the QGP medium, it is essential to examine the scattering cross-section of its constituent particles. The differential cross-section for the elastic scattering of the type (1+2 $\rightarrow$ 3+4) is given as
\begin{align}
	\frac{d \sigma}{dt}= \frac{1}{64\pi s }\frac{1}{p_{\rm cm}^2}\langle|\bm{\mathcal{M}}|\rangle_{12\rightarrow 34}^2~,
	\label{dsigmadt}
\end{align}
where $p_{\rm cm}$ is the momentum in the center of mass (COM) frame of incoming ($1,2$) and outgoing particles ($3,4$), which can be calculated as
\begin{align}
	p_\text{cm}=\frac{\sqrt{(s-(m_{1,3}-m_{2,4})^2)(s-(m_{1,3}+m_{2,4})^2)}}{2\sqrt{s}}~,
\end{align}
with $s$ being the Mandelstam variable. Note that in Eq.\eqref{dsigmadt}, the invariant matrix amplitude $\langle|\bm{\mathcal{M}}|\rangle_{12\rightarrow 34}^2$ is averaged over the initial and summed over the final spin states, and are computed perturbatively at the tree level for the elementary two body scattering process among the massive quasiparticle quarks/antiquarks and Gribov modified gluons for various possible channels ($s, t, u$), as shown in Fig.~[\ref{qqqq}] and Fig.~[\ref{qqbarqqbar}]. It is worth noting that even though the higher-order corrections were not considered in the evaluation of the scattering amplitude, those contributions are nevertheless not necessarily small. 

The utilization of different symmetries greatly simplifies the task when assessing the invariant scattering amplitudes involving quark-quark and quark-antiquark interactions. Below are a few examples:

\begin{itemize}
	\item $\langle |\bm{\mathcal{M}}|^2 \rangle_{d\bar{d} \rightarrow d\bar{d}}=  \langle |\bm{\mathcal{M}}|^2 \rangle_{u\bar{u} \rightarrow u\bar{u}}$; (charge symmetry).
	\item $\langle |\bm{\mathcal{M}}|^2 \rangle_{d\bar{u} \rightarrow d\bar{u}}= \langle |\bm{\mathcal{M}}|^2 \rangle_{u\bar{d} \rightarrow u\bar{d}}$; (charge conjugation).
	\item $\langle |\bm{\mathcal{M}}|^2 \rangle_{u\bar{u} \rightarrow d\bar{d}}(s,t)= \langle |\bm{\mathcal{M}}|^2 \rangle_{u\bar{d} \rightarrow u\bar{d}}(t,s)$; (crossing symmetry).
	\item $\langle |\bm{\mathcal{M}}|^2 \rangle_{uu \rightarrow uu}(u,t)= \langle |\bm{\mathcal{M}}|^2 \rangle_{u\bar{u} \rightarrow u\bar{u}}(s,t)$; (crossing symmetry).
\end{itemize}

The detailed formulation will be presented elsewhere. However, we note that in the limit $m_{i=1,2,3,4} \rightarrow 0$, our findings of scattering amplitudes are in line with the one presented in the Ref.~\cite{Cutler:1977qm}.

For the process of $qq(\bar{q})\rightarrow qq (\bar{q})$, the total scattering cross-section ($\sigma_{sc}$) can be expressed as
%The total scattering cross-section ($\sigma_{sc}$) for $ qq(\bar{q})\rightarrow qq (\bar{q})$  read as
%
\begin{align}
	\sigma_{sc}= \int_{t_-}^{t_+}dt \left(\frac{d\sigma}{dt}\right)[1 - f_3^0(T)][1 - f_4^0(T)]~,
\end{align}  
For $ ug\rightarrow ug $ process,  the total scattering cross-section can be written as
\begin{align}
	\sigma_{sc}= \int_{t_-}^{t_+}dt \left(\frac{d\sigma}{dt}\right)[1- f_3^0(T)][1 +{\rm{g}}_4^0(T)]~,
\end{align} 
Here, $(1 - f_{3,4}^0(T))$ and $(1 + {\rm{g}}_4^0(T))$ represent the Pauli blocking/Bose enhancement factors for fermions and bosons, respectively, these factors take into account the possibility that some of the final states are already occupied with the constituent particles. The functions ${\rm{g}}_i^0$ and $f_i^0$ represent the equilibrium distribution functions of the ``$i^{th}$'' particles of bosons and fermions, respectively, as defined in Eqs.~(\ref{g0}) and (\ref{f0}). 
The integration limit is fixed by considering the collision in the centre of mass frame, where the Mandelstam variable $t=-2p_\text{cm}^2(1-\cos \theta)$ with $-1 \le \cos \theta \le 1$. 

Here, all the partonic cross-sections are fixed as a function of temperature ($T$) and Mandelstam variables ($s, t, u$). 
\subsection{Findings of elastic cross-sections}
This subsection details the findings regarding the elastic cross-sections for quark-quark, quark-antiquark, and quark-gluon scattering.  
To provide a comprehensive comparison, we have illustrated each process using plots at two different temperature scales: at $T/T_c=1.2 $ and $T/T_c=2.2 $. It is important to note that we calculated the scattering cross-sections for these processes by using the quasiparticle masses of the quarks, as defined in Eq.~\eqref{quasimass}. Moreover, the interaction among the quarks in the QGP medium occurs via gluon exchange, following the Gribov prescription. We used the modified gluon propagator defined in Eq.~\eqref{eq:gribovprop} to account for this interaction.

The relevant $2 \to 2$ scattering processes contributing to the quark relaxation time include quark–quark, quark–antiquark, and quark–gluon scatterings. For the $u$ quark, these are $u u \to u u$ (t-, u-channel), $u d \to u d$ (t-channel), $u s \to u s$ (t-channel), $u \bar{u} \to u \bar{u}$ (s-, t-channel), $u \bar{u} \to d \bar{d}$ (s-channel), $u \bar{u} \to s \bar{s}$ (s-channel), $u \bar{d} \to u \bar{d}$ (t-channel), $u \bar{s} \to u \bar{s}$ (t-channel), $u g \to u g$ (t-channel), and $u \bar{u} \to g g$ (s-, t-, u-channel). The corresponding processes for $d$ and $s$ quarks are analogous.

%Here is the list of all possible scattering considered for the u$- $quark, along with their corresponding possible scattering channels:

%\begin{enumerate}
%	\item $ u + u \rightarrow u + u; {\hspace{0.2cm}} {\rm \textbf{channel/'s:}}\Rightarrow {\hspace{0.1cm}} t, u$~.
%	\item $u + d \rightarrow u + d; {\hspace{0.2cm}} {\rm \textbf{channel/'s:}}\Rightarrow {\hspace{0.1cm}} t$~. 
%	\item $u + s \rightarrow u + s; {\hspace{0.2cm}} {\rm \textbf{channel/'s:}}\Rightarrow {\hspace{0.1cm}} t$~.
%	\item $u + \bar{u} \rightarrow u + \bar{u}; {\hspace{0.2cm}} {\rm \textbf{channel/'s:}}\Rightarrow {\hspace{0.1cm}} s, t$~.
%	\item $u + \bar{u} \rightarrow d + \bar{d}; {\hspace{0.2cm}} {\rm \textbf{channel/'s:}}\Rightarrow {\hspace{0.1cm}} s$~.
%	\item $u + \bar{u} \rightarrow s + \bar{s}; {\hspace{0.2cm}} {\rm \textbf{channel/'s:}}\Rightarrow {\hspace{0.1cm}} s$
%	\item $u + \bar{d} \rightarrow u + \bar{d}; {\hspace{0.2cm}} {\rm \textbf{channel/'s:}}\Rightarrow {\hspace{0.1cm}} t$
%	\item $u + \bar{s} \rightarrow u + \bar{s}; {\hspace{0.2cm}} {\rm \textbf{channel/'s:}}\Rightarrow {\hspace{0.1cm}} t$
%	\item $u + g \rightarrow u + g; {\hspace{0.2cm}} {\rm \textbf{channel/'s:}}\Rightarrow {\hspace{0.1cm}} t$
%	\item $u + \bar{u} \rightarrow g + g; {\hspace{0.2cm}} {\rm \textbf{channel/'s:}}\Rightarrow {\hspace{0.1cm}} s,t,u$
%\end{enumerate}
%The same processes apply to both the $d$ and $s$ quarks.
%%%
\begin{figure*}
	\centering
	\includegraphics[width=8cm]{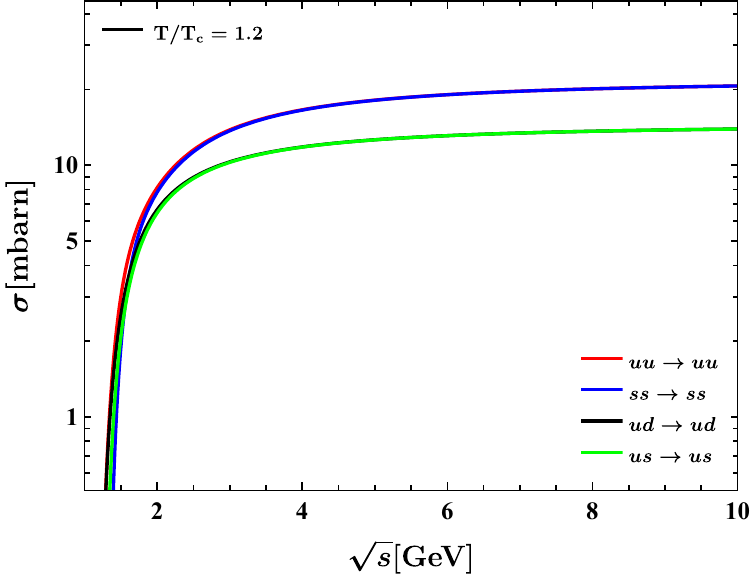}
	\hspace{1cm}
	\includegraphics[width=8cm]{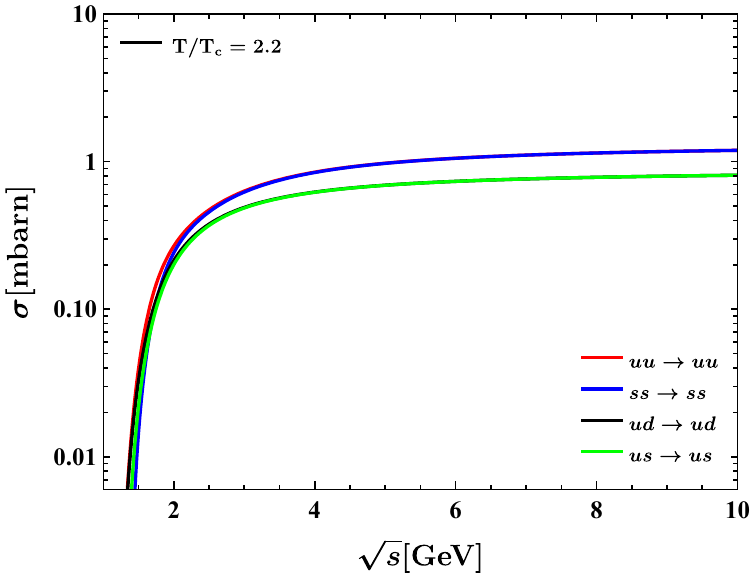}
	\caption{The cross-sections ($\sigma_{sc}$) for the quark-quark scattering are plotted as a function of $\sqrt{s}$, with the left plot corresponding to a scaled temperature of $T/T_c=1.2 $ and the right plot corresponding to $T/T_c=2.2 $.}
%		
%		The elastic cross-sections($\sigma_{sc}$) plotted against $\sqrt{s}$ for the quark-quark scattering. The left plot corresponds to the scaled temperature $T/T_c=1.2 $ while the right plot corresponds to the scaled temperature $T/T_c=2.2 $. } 
	\label{fig:qq}
\end{figure*}
Fig.~[\ref{fig:qq}] shows the cross-section for quark-quark scattering as a function of $\sqrt{s}$ at two different temperatures, $T/T_c=1.2 $ (left)  and $T/T_c=2.2 $ (right). %, as shown in  Fig~\ref{fig:qq}. %, we plot the cross-section for quark-quark scattering as a function of $\sqrt{s}$ at $T/T_c=1.2 $ (left)  and $T/T_c=2.2 $ (right).  
For a fixed temperature, the scattering cross-section increases with the center-of-mass energy $\sqrt{s}$, with a more rapid variation at lower $\sqrt{s}$. Specifically, the scattering cross-section increases sharply at smaller values of $\sqrt{s}$ and then tends to saturate, showing only a weak variation for $\sqrt{s}\gtrsim 2.5$ GeV. A comparison of the two panels further shows that the overall magnitude of the cross-section decreases as the temperature increases.
%Our analysis reveals that the scattering cross-section decreases as a function of $\sqrt{s}$ as temperature increases.
%as the scaled temperature increases from $T/T_c=1.2$ to $T/T_c=2.2$. 
%
Interestingly, we observe that the cross-sections are independent of the quasi-masses of the light and strange quarks, and remain the same for both $q+q \rightarrow q+q$ and $q+q^{\prime} \rightarrow q+q^{\prime}$ processes.
\begin{figure*}
	\centering
	\includegraphics[width=8cm]{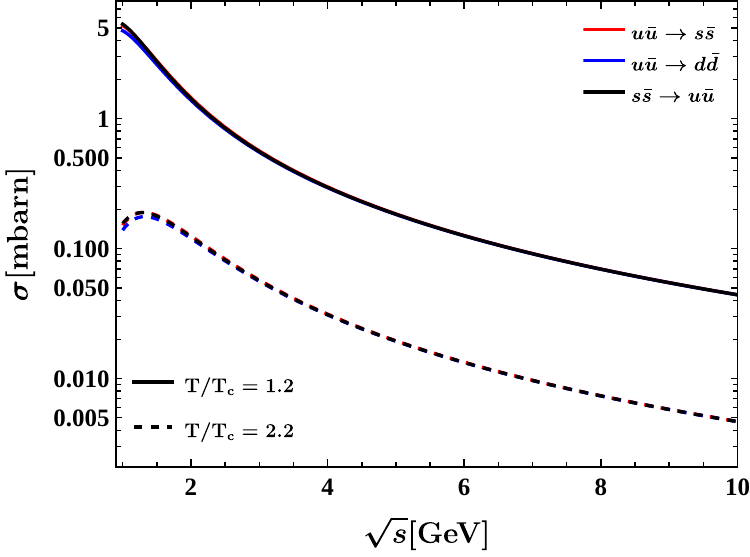}
	\hspace{10mm}
	\includegraphics[width=8cm]{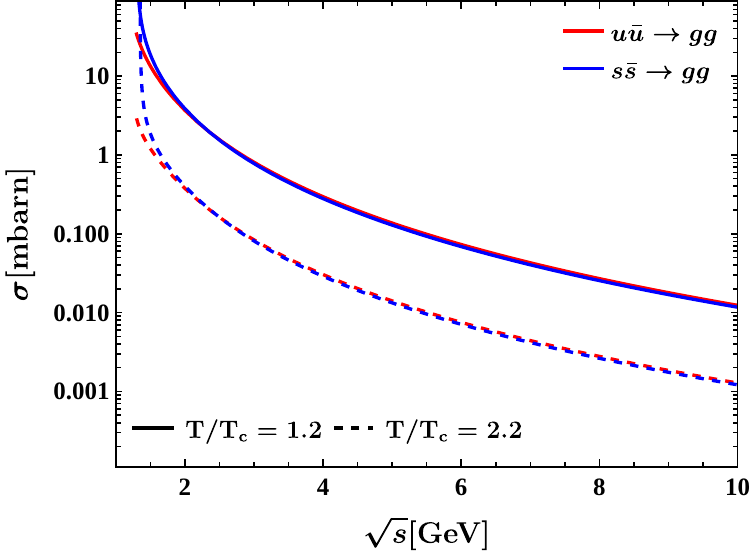}%[scale=0.36]{imr2.pdf}
	\caption{The scattering cross-sections for quark-antiquark (flavor changing) and quark-antiquark pair annihilation as a function of $\sqrt{s}$ at two different scaled temperatures, $T/T_c=1.2 $ (solid line) and $T/T_c=2.2 $ (dashed line)}.
	%	The scattering cross-sections of the quark-antiquark flavour-changing scattering are plotted against $\sqrt{s}$. The plot (on left) corresponds to the scaled temperature $T/T_c=1.2 $ while the plot(on right) corresponds to $T/T_c=2.2 $. } 
	\label{fig:qqbarflavourchange}
\end{figure*}

Fig.~[\ref{fig:qqbarflavourchange}] demonstrates how the cross-sections vary with the center-of-mass energy ($\sqrt{s}$) for two different processes: flavor-changing quark-antiquark scattering ($q+\overline{q} \rightarrow q^{\prime}+\overline{q}^{\prime}$) on the left, and the pair-annihilation of a quark-antiquark pair ($q+\overline{q}\rightarrow g+g$) on the right.
As shown in the figure, the scattering cross-section for both processes decreases monotonically as $\sqrt{s}$ increases. This behavior is in line with the predictions made by the dynamical quasiparticle model~\cite{Moreau:2019vhw}.
Interestingly, the difference in quasiparticle mass between light and strange quarks has no significant effect on the scattering cross-sections and remains the same for quark-antiquark scattering. Additionally, the difference in quasiparticle mass between light and strange quarks has minimal impact on the cross-sections of their respective pair-annihilations at smaller values of $ \sqrt{s} $. This observation underscores the consistent behavior in the scattering process, regardless of the quark flavors involved.
%
%\begin{figure}[H]
%	\centering
%	\includegraphics[scale=0.625]{uubargg&ssbargg}
%	\caption{The scattering cross-sections($\sigma_{sc}$) of quark-antiquark pair annihilation, plotted against $\sqrt{s}$. The solid line corresponds to the scaled temperature value $T/T_c=1.2 $ while the dashed line corresponds to $T/T_c=2.2 $.}
%	\label{fig:uubarggssbargg}
%\end{figure}
%
\begin{figure*}
	\centering
	\includegraphics[width=8cm]{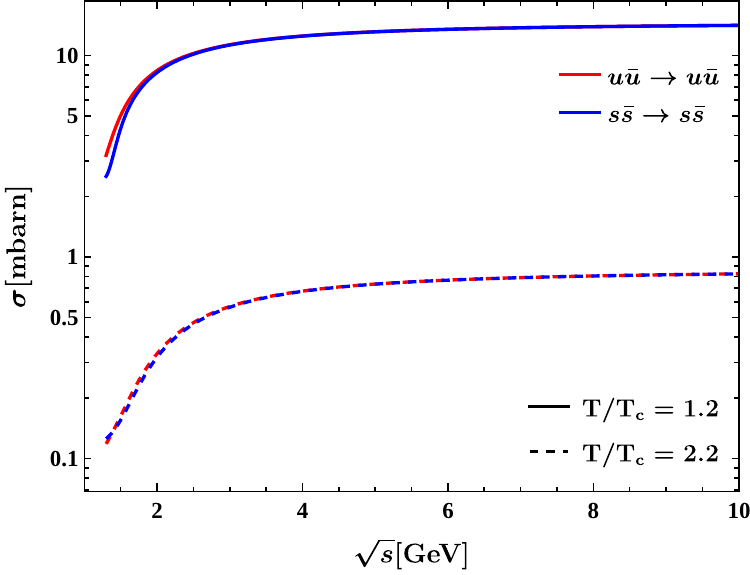}
	\hspace{10mm}
	\includegraphics[width=8cm]{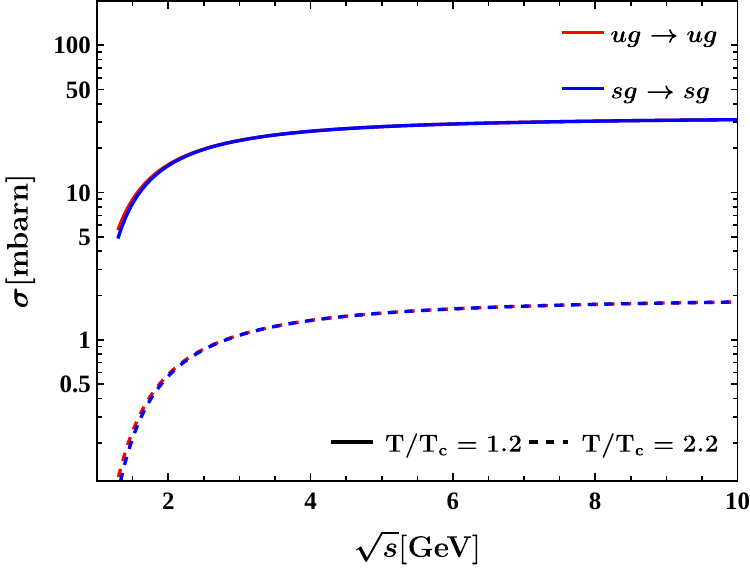}%[scale=0.36]{imr2.pdf}
	\caption{The cross-sections for the quark-antiquark (no flavor change) (left) and quark-gluon (right) scattering as a function $\sqrt{s}$. The solid line corresponds to  $T/T_c=1.2 $ and the dashed line corresponds to $T/T_c=2.2$}
	\label{fig:qqbar}
\end{figure*}

Figure~[\ref{fig:qqbar}] illustrates the behavior of the cross-section for quark-antiquark scattering ($q+\overline{q}\rightarrow q+\overline{q}$) (left) and quark-gluon scattering($q+g \rightarrow q+g$) (right) as a function of $\sqrt{s}$. Notably, for $\sqrt{s}\ge 3$~GeV, the scattering cross-section remains almost constant for both scaled temperature values, $T/T_c=1.2$ (solid line) and $T/T_c=2.2 $ (dashed line). As the temperature increases, the cross-section value decreases across the entire range of $\sqrt{s}$. This trend demonstrates a significant reduction in the scattering cross-section with increasing temperature. It reflects the impact of temperature variations on the dynamics of quark-antiquark pair interaction. 
%Note that in evaluating the scattering cross-section of the quark-gluon scattering, we have considered the $t-$channel scattering due to its main contribution.

The next step is to formulate the relaxation time ($\tau_R$), which is an essential parameter for calculating the transport coefficients of the QGP medium, which have been derived using the relaxation time approximation.
	
\section{Relaxation time $(\tau_R)$}\label{relaxationtime}
The transport properties of QGP are largely influenced by the relaxation time, $\tau_R$, which plays a crucial role in determining various transport coefficients. Therefore, accurately determining this parameter is of paramount importance in understanding the transport properties of QGP. In this study, we have used the method developed in~\cite{Sasaki:2008um} to calculate the relaxation time, $\tau_R$, based on the weighted thermally averaged quark-quark and quark-antiquark cross-sections. A similar approach has been previously used in literature to evaluate the relaxation time $\tau_R$~\cite{Soloveva:2020hpr, Mykhaylova:2019wci, Marty:2013ita, Sasaki:2008um, Danielewicz:1984ww}. The relaxation time for the species ``$i$'' is given by 
\begin{align}
	\tau_i^{-1}(T)= \sum_{j=q,\bar{q}, g}n_j(T)\overline{\sigma}_{ij}(s,T)~,
\end{align}
where $\overline{\sigma}_{ij}$ is the weighted thermal average of the cross-section. In the context of the scattering of the type $q\bar{q}\rightarrow q^{\prime}\bar{q}^{\prime}$, it is more convenient to define a general notation of the thermally averaged scattering cross-section, which is given as 
\begin{align}
	\overline{\sigma}_{12 \rightarrow 34}= \int_{\rm Th}^{\infty}ds \hspace{0.1cm} \sigma_{12 \rightarrow 34}(T,s) \mathcal{X}(T,s)~.
	\label{sigmabar}
\end{align}
Here, the threshold ${\rm Th=max}\{(m_1+m_2)^2,(m_3+m_4)^2\}$ and $\mathcal{X}(T,s)$ is the probability of finding the quark-quark and quark-antiquark pair with center-of-mass energy $\sqrt{s}$~\cite{Berrehrah:2013mua}
\begin{align}
	\mathcal{X}(T,s) =  \hspace{.1cm} \mathcal{C}\frac{E_1^\text{cm}(\sqrt{s}-E_1^\text{cm})}{\sqrt{s}}p_\text{cm}f^0_{i}(\sqrt{s}-E_1^\text{cm}) f^0_{i}(E_1^\text{cm}),
	\label{probability}
\end{align}
here $f^0_{i}(E_i)$ %=(1/{\rm Exp}(\beta E_i)+1)$ 
is the fermionic distribution function for quark-(anti) quark scattering as defined in Eq.~\eqref{f0} and $ E_1^\text{cm} $ is defined as
\begin{align}
	E_1^\text{cm}(s,T)=\frac{s-(m_1(T)^2-m_2(T)^2)}{2\sqrt{s}}.
\end{align}
Note that the normalization in Eq.~\eqref{probability} is fixed by 
\begin{align}
	[\mathcal{C}(T)]^{-1}= \int_{\rm Th}^{\infty} ds \hspace{0.1cm} \mathcal{X}(T,s)~.
\end{align}
Note that the electrical conductivity is primarily governed by the transport of electrically charged quarks. Since gluons are electrically neutral, they do not couple directly to the external electromagnetic field and therefore do not contribute independently to the electrical current. Their role enters indirectly through quark–gluon scattering processes, which affect the relaxation time of quarks. The relaxation time for the light quarks is obtained as~\cite{Soloveva:2020hpr}
\begin{align}
		\label{taul}
	\tau_u^{-1}(T)=& \hspace{.1cm} n_{\bar{u}}(\overline{\sigma}_{u\bar{u} \rightarrow u\bar{u}}+\overline{\sigma}_{u\bar{u} \rightarrow d\bar{d}}+\overline{\sigma}_{u\bar{u} \rightarrow s\bar{s}}+\overline{\sigma}_{u\bar{d} \rightarrow u\bar{d}}\nn \\ 
	&\hspace{-1cm}+\overline{\sigma}_{u\bar{u} \rightarrow gg}) 
	+ n_u(\overline{\sigma}_{uu \rightarrow uu}+\overline{\sigma}_{ud \rightarrow ud})+ n_s \overline{\sigma}_{us \rightarrow us} \nn\\ 
	&+n_{\bar{s}} \overline{\sigma}_{u\bar{s} \rightarrow u\bar{s}}+n_g \overline{\sigma}_{ug \rightarrow ug}~,
%	\label{taul}
\end{align}
and  for the strange quark is 
\begin{align}
		\label{taus}
	\tau_s^{-1}(T)=& \hspace{.1cm} 2 n_u\overline{\sigma}_{us \rightarrow us}+2n_{\bar{u}}\overline{\sigma}_{u\bar{s} \rightarrow u\bar{s}}+n_s \overline{\sigma}_{ss \rightarrow ss}+\\ \nn
	&n_{\bar{s}}( \overline{\sigma}_{s\bar{s} \rightarrow s\bar{s}}+\overline{\sigma}_{s\bar{s} \rightarrow gg}+2\overline{\sigma}_{s\bar{s} \rightarrow u\bar{u}})+n_g \overline{\sigma}_{sg \rightarrow sg} ~,
\end{align}
%where the degeneracy factor, $ d_i $, for quarks  and gluons is taken as $2\times N_c=6$ and  $2\times(  N_c^2-1)=16$, respectively.
Here $\overline{\sigma}_{12\rightarrow 34}$ is the weighted thermally averaged cross sections,  as defined in Eq.~\eqref{sigmabar}. In Eq.~\eqref{taul} and Eq.~\eqref{taus}, $n_i(T)$ is the equilibrium number density of the particles,  which is defined for quarks as
\begin{figure}
	\centering
		\includegraphics[scale=0.65]{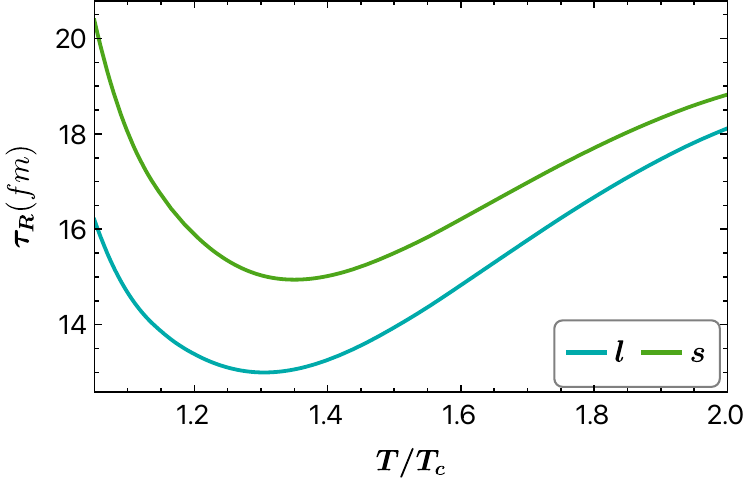}
	%	\hspace{10mm}
	%\includegraphics[scale=0.65]{numberdensity.pdf}%[scale=0.36]{imr2.pdf}
	\caption{The relaxation time ($\tau_R$) as a function of the scaled temperature %($T/T_c$). $T_c$ for the QGP medium has been taken to be 160 MeV. 
		for the light (Eq.~\ref{taul}) and  strange quarks (Eq.~\ref{taus}).  The light quark consists of $u$ and $d$ quarks.}
	\label{fig:tauR}
\end{figure} 
\begin{equation}
	n_i(T)=d_i\int\frac{d^3p_i}{(2\pi)^3}f_i^0~,
%	\label{nd}
\end{equation}
and for gluons as
\begin{equation}
	n_g(T)=d_g\int \frac{d^3p}{(2\pi)^3}{\rm{g}}^0~.
%	\label{nd}
\end{equation}
Here, $f_i^0$ is the equilibrium distribution function for quarks as defined in Eq.~\eqref{f0}, while ${\rm{g}}^0$ is the equilibrium distribution function for the Gribov modified gluons, with ${\rm{g}}_\pm^0$ corresponding to the two dispersion branches as mentioned in Eq.~\eqref{g0}. In Fig.~[\ref{fig:tauR}], we present the behavior of relaxation times as a function of scaled temperature. It is measured at vanishing chemical potential. It is observed that the relaxation time for both light and strange quarks first decreases and then increases with temperature.
%%Here $g^0$ denotes the equilibrium distribution for the Gribov plasma, $.
\begin{figure}
	\centering
%	\includegraphics[scale=0.65]{tau.pdf}
%	\hspace{10mm}
	\includegraphics[scale=0.65]{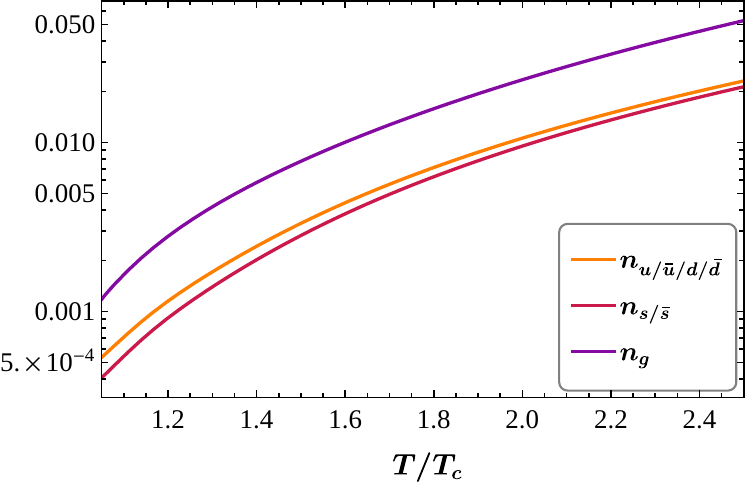}%[scale=0.36]{imr2.pdf}
	\caption{The equilibrium number density of the light and strange quark along with Gribov modified gluons as a function of $T/T_c$ in the range $1.05 \le T/T_c \le 2.5$. }
	\label{fig:numberdensity}
\end{figure}
In Fig.~[\ref{fig:numberdensity}], the variation of the number densities of the light quarks and the strange quark is plotted against the scaled temperature $T/T_c$. We see that the number density of the Gribov modified gluons is significantly higher compared to quasiparticle quarks for the entire range of scaled temperature.
%%%%%%%%%%%%%%%%	
\section{Electrical conductivity}\label{electricalconductivity}
		In this section, we have computed the electrical conductivity of the QGP medium, which quantifies the ability of a system to conduct electric charges. After solving the relativistic Boltzmann kinetic equation in the relaxation time approximation (RTA), one can obtain the electrical conductivity expression as~\cite{Thakur:2017hfc, Mykhaylova:2020pfk}
	\begin{align}
		\sigma_\text{el}=\frac{1}{3 T} \sum_{i = q, \bar{q}} \int \frac{d^3 k}{(2 \pi)^3} \frac{k^2}{E_i^2} q_i^2 d_i \tau_i f_i^0\left(1-f_i^0\right),
		\label{eq:eleccond}
	\end{align}
    where $\tau_i$ is the relaxation time of the $i$-th quark or antiquark species, defined in Section~\ref{relaxationtime}.
	The electric charge $q_i$ for u, d and s quarks are $q_u=-q_{\bar{u}}=2e/3$ and $q_{d,s}=-q_{\bar{d},\bar{s}}=-e/3$, respectively. The electron charge $e=\sqrt{4\pi \alpha}$ with the fine structure constant, $\alpha=1/137$ and $f_i^{0}$ is the equilibrium distribution, which is defined in Eq.~\eqref{f0}. %$ f^{0} $ and gluon, $ b^{0} $ at $ \mu=0 $ 
%%%%%%%%%%%%%%%%
\begin{figure}
\includegraphics[scale=0.525]{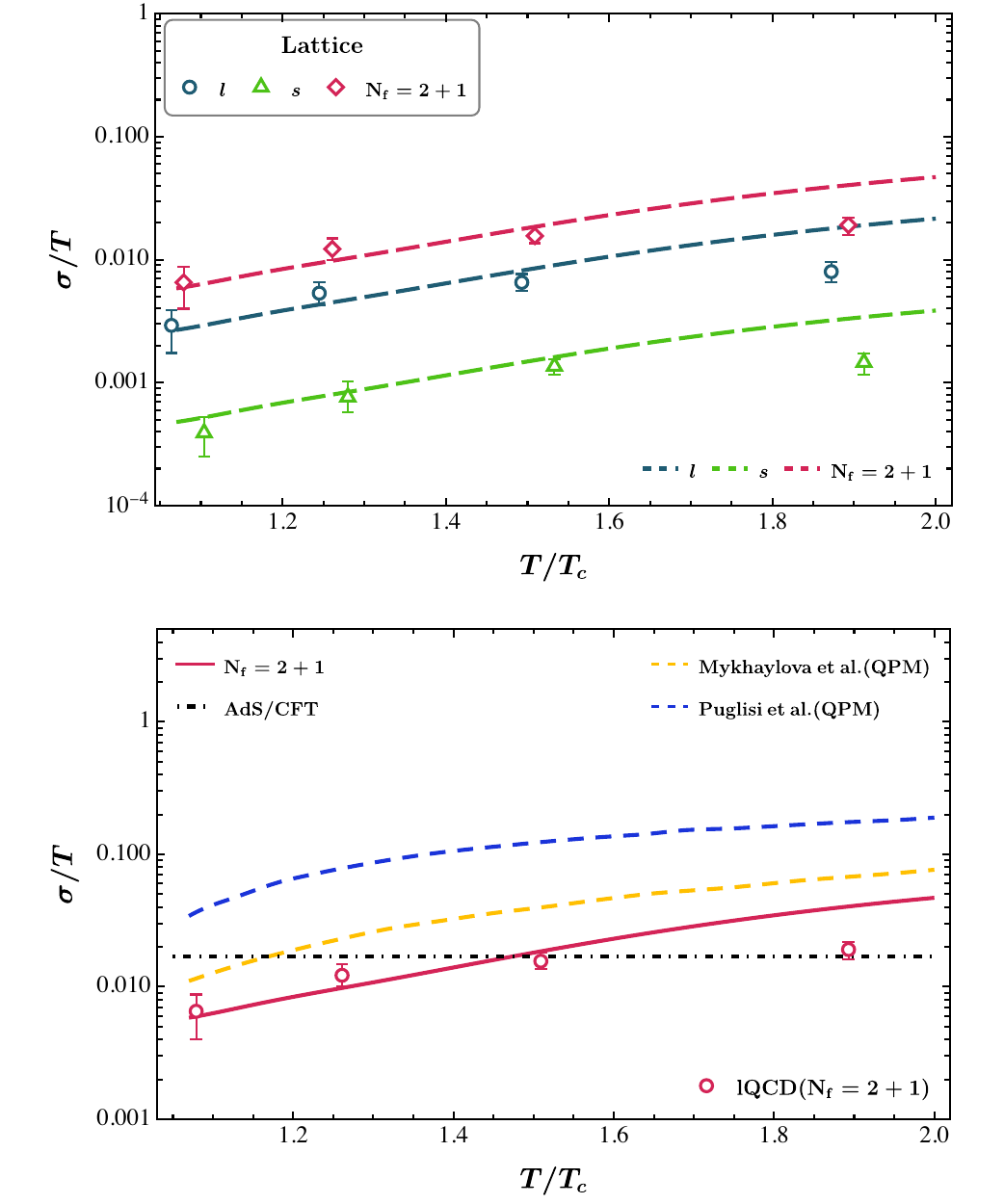}
\caption{The scaled electrical conductivity as a function of the scaled temperature for different flavors ($l$, $s$, and $N_f=(2+1)$) (left) and for $N_f=(2+1)$ (right). The blue, green, and red symbols from lattice data~\cite{Aarts:2014nba}, blue and yellow lines from QPM~\cite{Puglisi:2014pda, Mykhaylova:2020pfk},  and the black dot-dashed line from the AdS/CFT approach~\cite{Caron-Huot:2006pee}.}
\label{fig:sigmabyt}
\end{figure}
%%%
In Fig.~[\ref{fig:sigmabyt}], we present a plot of the electrical conductivity as a function of scaled temperature. The plot on top compares our electrical conductivity results for different flavors ($l$, $s$, and $N_f=(2+1)$) with the lattice data~\cite{Aarts:2014nba}. The red, blue, and green dashed lines correspond to the light (consisting of $u$ and $d$ quarks), strange, and $N_f=(2+1)$ flavors, respectively, along with the same color symbols taken from the lattice data. 
In plot on the bottom, we show the variation of the electrical conductivity for $N_f=(2+1)$ (solid red line) compared with the results from the Quasiparticle Model (QPM) (blue and yellow dashed lines)~\cite{Puglisi:2014pda, Mykhaylova:2020pfk}, lattice QCD (red circle), and the black dot-dashed line represents the value of $\sigma/T=e^2N_c^2/16\pi\sim0.017$ obtained using the AdS/CFT approach~\cite{Caron-Huot:2006pee}. Our results are in close agreement with the lattice data. It is worth mentioning that the antiquark contribution to the electrical conductivity ($\sigma_{u,d,s}$) has been neglected while comparing our findings with those of the lattice calculation of the same, considering that the lattice findings do not include the contribution from respective antiquarks, and are just for quarks. 

%%%%%%%%%%%%%%%%%%%%%%%%%%%%%%%%%%%%%%%%%%%%%%%%%
\section{Summary and outlook}\label{summary}
In this work, we have investigated the electrical conductivity scaled with temperature $T$ for the medium QGP, consisting of light ($u$ and $d$) and strange quarks, using the quasiparticle approach within the kinetic theory framework in relaxation time approximation. 
%The exchanged gluons between quarks have been modified by the Gribov idea, where the Gribov parameter has been fixed using lattice data of thermodynamic quantities. 
The exchanged gluons between quarks are modified by the Gribov prescription, where the temperature dependence of the Gribov parameter is fixed from pure gauge lattice thermodynamics. Also, the quasi-masses of the light and strange quarks are parameterised with the running coupling $g(T)$, which is obtained using the lattice equation of state of $(2+1)$-flavor QCD.

The relaxation time $\tau_R$ has been an important dynamical parameter in determining the transport coefficients, including the electrical conductivity. In this work, the relaxation time has been evaluated using the thermally averaged cross-section of all possible quark-(anti)quark scatterings to the lowest order; however, it is worth noting that higher-order corrections may have a significant impact on the outcome.  

We also investigated the quark flavor dependence of the electrical conductivity and compared it with the available lattice findings. We observe a good match with the lattice data, particularly near the phase transition temperature. We also compared our final result with the previous finding of the electrical conductivity based on the quasiparticle model.

\section*{Acknowledgements}
Discussion with Aritra Bandhopadhyay and Hiranmay Mishra is highly appreciated. L.~T.~ is supported by the National Research Foundation (NRF) funded by the Ministry of Science of Korea (Grant No. 2021R1F1A1061387). N.H. is supported in part by the SERB-MATRICS under Grant No. MTR/2021/000939
	% color red bracket
	
	%%%%%%%%%%%%%%%%%%%%%%%%%%%%%%%%%%%%%%%%%%%%%%%%%

\end{document}